\documentclass[journal,comsoc]{IEEEtran}

\usepackage[T1]{fontenc}

\ifCLASSINFOpdf
\else
\fi

\usepackage{amsmath}
\usepackage{comment} %
\usepackage{cite}
\usepackage{graphicx}
\usepackage{epsfig,graphics,subfigure,psfrag,amsmath,amssymb}
\usepackage{booktabs}
\usepackage{amsfonts}
\usepackage{epstopdf}
\usepackage{amssymb}
\usepackage{array}
\usepackage{amsmath}
\usepackage{makecell}
\usepackage{multirow}
\usepackage{diagbox}

\begin{document}
%
\title{AI/ML for Beam Management in 5G-Advanced: A Standardization Perspective}

\author{Qing~Xue,~\IEEEmembership{Member,~IEEE,}
	Jiajia~Guo,~\IEEEmembership{Member,~IEEE,}
    Binggui~Zhou,
	Yongjun~Xu,~\IEEEmembership{Senior Member,~IEEE,}
	Zhidu~Li,~\IEEEmembership{Member,~IEEE,}
	and Shaodan~Ma,~\IEEEmembership{Senior Member,~IEEE}

	\thanks{Qing Xue, Yongjun Xu and Zhidu Li are with Chongqing University of Posts and Telecommunications, Chongqing 400065, China.}
	
	\thanks{Jiajia Guo, Binggui Zhou and Shaodan Ma are with University of Macau, Macao SAR, China.}

}

\maketitle

\begin{abstract}
In beamformed wireless cellular systems such as 5G New Radio (NR) networks, beam management (BM) is a crucial operation. In the second phase of 5G NR standardization, known as 5G-Advanced, which is being vigorously promoted, the key component is the use of artificial intelligence (AI) based on machine learning (ML) techniques. AI/ML for BM is selected as a representative use case. This article provides an overview of the AI/ML for BM in 5G-Advanced. The legacy non-AI and prime AI-enabled BM frameworks are first introduced and compared. Then, the main scope of AI/ML for BM is presented, including improving accuracy, reducing overhead and latency. Finally, the key challenges and open issues in the standardization of AI/ML for BM are discussed, especially the design of new protocols for AI-enabled BM. This article provides a guideline for the study of AI/ML-based BM standardization.
\end{abstract}

\IEEEpeerreviewmaketitle

\section{Introduction}
The evolution of 5G New Radio (NR) has progressed swiftly since the 3rd generation partnership project (3GPP) standardized the first NR release (Release 15) in mid-2018, which laid the basis for the commercial deployment of 5G worldwide. The millimeter wave (mmWave) spectrum is recognized as a crucial enabler in fulfilling the performance demands of NR, offering vast available bandwidth and unparalleled data rates. However, this advantage is tempered by the fact that mmWave signals undergo significant propagation attenuation. To deal with the impairments, the 5G Node Base Station (gNB) and User Equipments (UEs) must form highly directional transmission links to secure an adequate link budget and maintain satisfactory communication quality. This necessitates precise alignment of transmitter and receiver beams through a process known as \emph{beam management (BM)} \cite{Tutorial-Beam-Management}. Although 3GPP Release 15 supports flexible BM functionality to accommodate different implementation and usage scenarios, for UEs with moderate to high mobility speeds, the signaling overhead and latency associated with BM emerge as critical bottlenecks. In response to this issue, various enhancements for BM in 5G at mmWave frequencies are specified in Releases 16 and 17.

A pivotal milestone in the progression of 5G mobile networks toward 6G and beyond is designated as 5G-Advanced \cite{Standardization-5G-Advanced}. This term, officially sanctioned by 3GPP in April 2021, marks a significant evolutionary step. A cornerstone of 5G-Advanced is the integration of artificial intelligence (AI), rooted in machine learning (ML) methods \cite{Pervasive-Machine-Learning}, which facilitates the delivery of smart network solutions. This advancement extends the network's capabilities to a myriad of novel use cases that surpass the scope of previously established use cases and deployment scenarios. As the beginning of 5G-Advanced, 3GPP Release 18 considers AI-native air interface \cite{3GPP-Artificial-Intelligence}. Using a use-case-centric approach, the study item focuses on three use cases: channel state information feedback enhancement, BM, and positioning accuracy enhancements \cite{3GPP-RP-213599}. This article is dedicated to the study of the second use case. AI/ML algorithms can be employed for beam prediction to reduce overhead and latency and improve prediction accuracy. Over the past few years, the academic community has made remarkable achievements across various application scenarios. For a comprehensive overview, one can refer to \cite{xue2023survey}. Unlike the academic overview provided in \cite{xue2023survey}, this article delves into the application of AI/ML for BM from an industrial perspective. It explores the attendant challenges and unresolved issues associated with deployment and emphasizes the significance of these developments in the context of 5G-Advanced standardization.

Traditional BM techniques may struggle to keep up with the increasing complexity and dynamic nature of modern communication systems. AI offers the ability to handle this complexity by learning from data and adapting to changing conditions in real-time. This can lead to improved performance, increased efficiency, and reduced costs. Additionally, AI can help in optimizing network resources, enhancing the user experience, and enabling new capabilities that may not be possible with traditional methods. As the demand for high-speed, high-quality, and reliable communication continues to grow, integrating AI into BM becomes increasingly important to meet these challenges. Beyond academic research, exploring AI for BM from an industrial perspective is crucial. It addresses the need for interoperability among diverse systems, ensuring they can work together seamlessly. It enhances development efficiency, setting a reliable foundation for the technology. Standardization also promotes innovation by providing a common ground for building advanced features. It aligns with regulatory requirements, facilitating market entry. Furthermore, it ensures user trust by delivering consistent performance and fosters cost-effective solutions by streamlining processes and encouraging global collaboration in technological advancement.

\begin{figure*}[t]
	\centering
	\includegraphics[width=17cm]{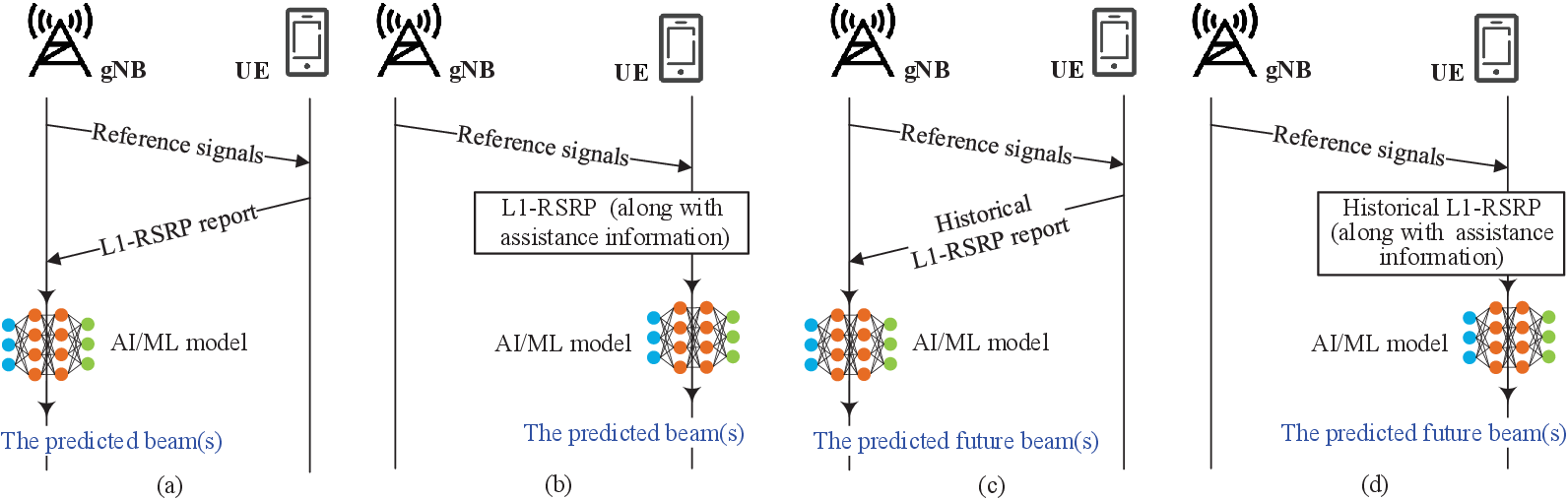}
	\caption{Typical BM procedures for (a) SBP with a gNB-side AI/ML model, (b) SBP with a UE-side AI/ML model, (c) TBP with a gNB-side AI/ML model, and (d) TBP with a UE-side AI/ML model.}
	\label{fig:BM}
\end{figure*}

\section{Representative BM Frameworks}

\subsection{Non-AI BM Frameworks}
Communications at high frequencies require directionality to compensate for severe propagation loss. Therefore, UEs and gNBs in 3GPP NR need to align their beams during both initial access and data transmissions to ensure sufficient beamforming gain. Toward this goal, 3GPP NR has specified a set of basic Layer 1 (L1)/ Layer 2 procedures under the term BM, which include at least the following aspects:

\emph{1) Beam sweeping}, i.e., beams transmitted and/or received in a predetermined manner during a time interval to cover a spatial area. In general, this process is carried out with an exhaustive search, where the gNB/UE has a predefined codebook of directions covering the angular space for transmitting/receiving synchronization and reference signals (RSs) sequentially.

\emph{2) Beam measurement}, i.e., the gNB (for uplink) or UE (for downlink) measures the characteristics/quality of the received signals. Typically, signal quality is expressed in terms of reference signal received power (RSRP) or signal-to-interference-plus-noise ratio (SINR).

\emph{3) Beam determination}, i.e., the gNB or UE selects their own suitable transmit/receive beam(s) according to the results obtained during the beam measurement procedure. The selected beam(s) that usually experienced the maximum RSRP or SINR will be used to establish directional communication for the subsequent transmission and reception.

\emph{4) Beam reporting}, i.e., the procedure with which the UE reports information about the beam quality and beam decision to the radio access network.

These processes are cyclically executed to continuously refresh and maintain the optimal pairing of transmit-receive beams as time progresses. For idle UEs, BM is the basis for designing a directional initial access strategy. For connected UEs, the directional path may deteriorate rapidly due to the dynamics of mmWave channels, requiring BM to maintain beam alignment, which is called beam tracking. Built on Release 15, 3GPP Releases 16 and 17 enhanced BM in various design aspects. For example, L1-RSRP-based beam measurement is supported in Release 15, while Release 16 ushers in the use of L1-SINR for beam measurement, enhancing the process of beam selection with keen awareness of interference. As downlink and uplink beam indications share similar characteristics, Release 17 specifies a unified transmission configuration indicator (TCI)-based framework for downlink and uplink beam indication to reduce signaling overhead and latency \cite{Massive-MIMO}. In particular, Releases 16 and 17 enhance the operation of multiple transmission/reception points (TRPs), while Release 15 focuses on single TRP-based operation.

Despite the aforementioned advancements, traditional BM often encounters substantial overhead, augmented communication delays, and diminished spectral efficiency. This situation arises primarily because a considerable quantity of beam measurements and reporting is necessary to accurately identify the most effective beam pair between a gNB and a UE. The situation is even worse when UE mobility, especially high mobility, is considered. In high-speed scenarios, it is almost impossible to maintain a desired narrow beam pair due to the latency of the BM procedure. To address these issues, AI/ML-based solutions have attracted great interest, especially for beam prediction and beam measurement feedback compression.

\subsection{AI-enabled BM Frameworks}
Among the array of AI/ML-based frameworks, 5G-Advanced focuses on those that necessitate little or no adjustments for integration with current systems. On the other hand, methodologies that profoundly overhaul existing systems are anticipated to be the subject of research in 6G and subsequent generations. In line with the 3GPP design for 5G-Advanced \cite{3GPP-TR-38843,3GPP-RAN1-113-Discussion}, we discuss two representative BM use cases based on AI/ML in the following and consider a downlink architecture. For ease of explanation, we define two beam sets, i.e., $\cal A$ and $\cal B$, where $\cal A$ is for beam prediction and $\cal B$ is for beam measurement.

{\bf{Spatial Beam Prediction (SBP):}} In this framework, we focus on spatial downlink beam prediction for $\cal A$ utilizing the measurements from $\cal B$. Within this schema, the training and inference processes of the AI/ML model can be executed either at the gNB end or on the UE side, offering flexible deployment options to optimize network performance.

\emph{Option 1: gNB-side SBP.} Fig.~\ref{fig:BM}(a) illustrates a typical procedure for SBP utilizing an AI/ML model located at the gNB side. In this scenario, the measured L1-RSRP values from $\cal B$  serve as inputs for the AI/ML model, enabling it to accurately predict the optimal beams for $\cal A$. In general, for AI/ML output used for beam prediction, transmit and/or receive beam identifier (ID) is preferred. In this subcase, there are two alternatives for the relationship between $\cal A$ and $\cal B$. One is that $\cal B$ is a subset of $\cal A$. The benefit is the reduction in overhead and latency, since gNB will send fewer RSs in comparison with legacy and only needs to scan a subset of beams for transmit beam sweeping. The other alternative is that $\cal B$ consists of wide beams while $\cal A$ consists of narrow beams. In the context of wide-to-narrow beam prediction, the measurement of wide beams facilitates the anticipation of one or a cluster of narrow beams, thereby obviating the need for direct measurements of these narrower beams. Hence, the overhead and latency can also be reduced.

\emph{Option 2: UE-side SBP.} Fig.~\ref{fig:BM}(b) shows a typical procedure for SBP with a UE-side AI/ML model, where UE measures the beams from $\cal B$ and reports the strongest beam index(es) optionally along with the corresponding predicted L1-RSRP(s) within $\cal A$. In this scenario, the correlation between $\cal A$ and $\cal B$ mirrors that which exists in the context of a gNB-side AI/ML model deployment. Similarly, the merits of this subcase align with those identified in option 1. In particular, some assistance information that provides spatial association between $\cal A$ and $\cal B$ may be used as an auxiliary input to on-device AI/ML models for beam prediction. For example, having information about the gNB beam shape, e.g., knowing the beam boresight directions (azimuth and elevation) along with 3dB beamwidth for the beams from $\cal A$ and $\cal B$, could enhance the prediction accuracy. Meanwhile, a richer UE report could be achieved, which contains information about gNB beams from $\cal A$ with higher angular resolution.

{\bf{Temporal Beam Prediction (TBP):}} The framework allows for the anticipation of temporal downlink beams for $\cal A$, leveraging historical measurement data from $\cal B$. In parallel with SBP, both the gNB and UE have the discretionary ability to undertake AI/ML model training within this system.

\emph{Option 1: gNB-side TBP.} Fig.~\ref{fig:BM}(c) shows a typical procedure for TBP, utilizing a gNB-side AI/ML model. Here, the gNB forecasts the optimal beams within $\cal A$ for subsequent $y$ BM cycles. This prediction is based on L1-RSRP measurements for $\cal B$, gathered from $x$ consecutive cycles. These measurements serve as inputs to the AI/ML model and are part of a recurring measurement sequence spanning every $x+y$ contiguous cycles. In addition to the two alternatives in SBP, there is a third alternative relationship between $\cal A$ and $\cal B$ in TBP, i.e., $\cal A$ and $\cal B$ are the same. This can be viewed as a purely temporal prediction without any spatial prediction. One benefit of the TBP in the gNB-side subcase is conducive to the link adaptation of gNB. For example, with the knowledge of the predicted future beam, the gNB can perform beam switching or adjust modulation and coding schemes in advance to avoid potential link failures or retransmissions.

\emph{Option 2: UE-side TBP.} Fig.~\ref{fig:BM}(d) depicts a conventional method for TBP employing a UE-side AI/ML model, where UE predicts the L1-RSRP(s) for several future time instances based on the previous measurements of $\cal B$. The benefit of this subcase is similar to that of the gNB-side subcase. Meanwhile, as in the option 2 for SBP, to enable inference of on-device AI/ML models for beam prediction, the availability of some assistance information from gNB (e.g., gNB beam boresight directions, 3dB beamwidth or beam shape of gNB beams) at the UE side is assumed. As an illustrative example, if the relative direction of the gNB beams is known, given the history of previous beam measurements, the UE may perform a more informed prediction compared to the scenarios without this information. Indeed, beyond the previously mentioned beam shape-related information, assistance information could encompass elements such as the gNB codebook ID, gNB antenna configuration ID, and information pertinent to positioning.

\begin{table}[t]
	\centering
	\caption{Comparison of evaluation results between non-AI and AI-enabled BM.} \label{Evaluation}
		\begin{tabular}{|c|cc|c|c|}
			\Xhline{0.5pt}
			\Xhline{0.5pt}
			\multirow{3}{*}{\textbf{Source}} & \multicolumn{2}{c|}{\multirow{3}{*}{\textbf{Framework}}}  & \multicolumn{2}{c|}{\textbf{KPI}} \\ \cline{4-5}
& & &\textbf{{\makecell[c]{Prediction \\Accuracy}}} 	&\textbf{{\makecell[c]{L1-RSRP\\ Difference}}}	\\ \hline

\multirow{4}{*}{Qualcomm \cite{Evaluations-Qualcomm-2305330}} &\multirow{2}{*}{SBP} &AI-enabled &63.50\% &0.36 dB \\ \cline{3-5}
 & &Non-AI &10.70\% &4.22 dB\\ \cline{2-5}
 &\multirow{2}{*}{TBP} &AI-enabled &81.60\% &0.55 dB\\ \cline{3-5}
 & &Non-AI &78.28\% &0.84 dB \\ \hline

\multirow{4}{*}{\makecell[c]{Huawei \cite{Evaluations-Huawei-2304655}}} &\multirow{2}{*}{SBP} &AI-enabled &75.40\% &-0.60 dB\\ \cline{3-5}
& &Non-AI &29.40\% &-1.54 dB\\ \cline{2-5}
&\multirow{2}{*}{TBP} &AI-enabled &56.35\% &-2.96 dB\\ \cline{3-5}
& &Non-AI &63.25\% &-2.14 dB\\
			\Xhline{0.5pt}
			\Xhline{0.5pt}
		\end{tabular}
\end{table}

For AI/ML-based BM performance evaluation, we consider two common key performance indicators (KPIs), namely, beam prediction accuracy and average L1-RSRP difference of the Top-1 predicted beam. Some evaluation results of non-AI and AI-enabled downlink transmit beam prediction are presented in Table~\ref{Evaluation}, where the simulation details can be found in \cite{Evaluations-Qualcomm-2305330,Evaluations-Huawei-2304655}. As anticipated, in SBP scenarios, AI-powered methods significantly exceed traditional non-AI strategies in achieving higher beam prediction accuracy and reducing the average L1-RSRP difference. However, for certain TBP instances, it has been observed that the Top-1 prediction accuracy of AI-facilitated approaches occasionally falls short when compared to non-AI methods. This observation underscores the value of adopting Top-$N$ ($N>1$) prediction to complement the singular focus on Top-1 inference when utilizing AI/ML models. The AI/ML-based Top-$N$ inference can achieve significant gain over Top-1 inference with only a slight increase in overhead. Further, the larger the value of $N$ chosen, the better the prediction performance of the AI/ML-based method. Employing different AI/ML technologies and taking different non-AI methods as benchmarks, the specific simulation results are usually different, but roughly the same evaluation conclusions can be drawn.

\begin{table*}
	\caption{Comparison between non-AI and AI-enabled BM frameworks.}
	\label{Framework}
	\centering
	\begin{tabular}{|c|c|c|c|c|}
		\hline
		&Non-AI BM &AI-enabled BM \\
		\hline
		{\makecell[l]{Time\\Complexity}} &{\makecell[lp{7.5cm}]{Higher, due to the need for exhaustive searches or grid-based methods, which can be time-consuming in complex environments.}} &{\makecell[lp{7.5cm}]{Lower, as AI learns from data and adapts quickly to new scenarios.}} \\
		\hline
		{\makecell[l]{ Computational\\Complexity}}&{\makecell[lp{7.5cm}]{Higher, with large-scale systems or numerous variables, requiring significant computational resources to process and analyze data.}} &{\makecell[lp{7.5cm}]{Lower, as AI automates optimization and can approximate complex functions.}} \\
		\hline
		{\makecell[l]{Model\\Development}}&{\makecell[lp{7.5cm}]{Less intensive, as traditional methods do not require training on large datasets.}} &{\makecell[lp{7.5cm}]{Intensive, requiring large datasets and computational resources for training.}} \\
		\hline
		{\makecell[l]{Scalability}}&{\makecell[lp{7.5cm}]{More challenging, due to the potential combinatorial explosion in the number of beam configurations with increasing system size.}} &{\makecell[lp{7.5cm}]{Easier, as AI can be parallelized and distributed across multiple processing units.}} \\
		\hline
		{\makecell[l]{Adaptability}}&{\makecell[lp{7.5cm}]{Low, as traditional methods are less flexible and may require manual adjustments or recalculations for new scenarios.}} &{\makecell[lp{7.5cm}]{High, AI can learn and improve over time, adapting to new environments and scenarios.}} \\
		\hline
	\end{tabular}		
\end{table*}

\subsection{Framework Comparison}
In the BM procedure, the propagation environment serves as the pivotal element that defines the distribution of beams, and this can essentially be considered as a form of environmental knowledge. AI/ML can precisely learn beam characteristics and distribution from data samples, thus achieving high beam prediction accuracy. In contrast, non-AI methods, which design beam prediction strategies based on an assumed distribution, often do not match practical systems well, resulting in poor performance. Table~\ref{Framework} shows a comparison of the two types of BM frameworks in several aspects.

Among the two AI-enabled BM frameworks, SBP focuses on optimizing beam direction in real-time by leveraging spatial data and environmental characteristics, aiming to enhance signal strength and system efficiency. In contrast, TBP utilizes temporal data to anticipate future beam orientations, allowing for proactive adjustments to maintain communication quality as conditions evolve. While SBP focuses on immediate enhancements, TBP aims to reduce latency and improve user experience over time, with each facing unique challenges related to data accuracy and environmental dynamics.

UE-sided SBP and TBP models require lower complexity due to the limited computational resources and power constraints of user devices, necessitating simpler algorithms. In contrast, gNB-sided models have access to more powerful processing capabilities, enabling the use of more complex and potentially more accurate AI models. However, gNB-side models must also consider factors like latency, scalability, and real-time processing, which can influence their complexity. Striking a balance between accuracy and resource efficiency is key for effective beam prediction on both sides.

\section{Standardization of AI/ML for BM}
In practice, transceiver designs from different companies can employ diverse technologies without needing to be shared. However, there needs to be a consensus on standards for implementing AI/ML in BM to ensure compatibility and interoperability across different network vendors and operators. Many issues should be explored and clarified in 5G-Advanced. On the basis of the scope and objectives of our article, this section focuses more on standardization rather than discussing detailed techniques.

\subsection{Scope of AI/ML for BM}
\emph{1) Accuracy Improvement:} The downlink performance depends heavily on the link quality corresponding to the downlink transmit and/or receive beam or the beam pair. However, the beam prediction accuracy in practical systems is difficult to meet future requirements. Therefore, it is necessary to develop an AI-enabled method that can significantly improve prediction accuracy without increasing overhead.

\emph{2) Overhead Reduction:} The upper bound of BM performance is to employ exhaustive beam sweeping over all available beams. However, the legacy method would bring heavy measurement/RS overhead and power consumption, which has a fatal impact on mobile scenarios. Employing AI/ML for sparse beam sweeping and beam prediction is a promising means of potentially saving the signaling overhead while providing large latency gains.

\emph{3) Latency Reduction:} For cases when UE and gBS have a large number of beams, performing BM procedures sequentially results in large measurement and reporting latency. If AI/ML-based solutions are considered, the best beam pair link can be predicted from a subset of channel or RSRP measurements thereby reducing the latency of beam acquisition or tracking without compromising prediction accuracy.

\emph{4) Others:} BM is closely linked to how resources are allocated. AI/ML can optimize the allocation of frequency channels and time slots in conjunction with beam selections to maximize network efficiency. AI/ML algorithms can predict and mitigate interference between overlapping beams to improve the overall signal quality and network performance. AI can detect and diagnose problems in BM, such as misalignment or signal degradation, enabling rapid response to maintain service quality. AI/ML can optimize BM operations to reduce the power consumption of network devices, contributing to greener wireless networks.

\subsection{Open Issues}
\emph{1) Complexity of AI-enabled BM:} While AI-enhanced BM is anticipated to excel beyond traditional algorithms, its computational demands pose a significant challenge. On one hand, the efficacy of AI-based techniques is often augmented through the implementation of numerous neural network (NN) layers or by expanding the width of these NN layers. In other words, the performance gains are purchased by significantly increasing the complexity of the AI/ML model. On the other hand, the NN weights must be stored,  thereby consuming storage resources of UE. Thus, the complexity of deploying AI-enabled BM should be assessed and reduced to balance costs and benefits. Typically, the number of floating-point operations, memory access cost, and parallelism degree must be considered in the NN inference speed evaluation \cite{AI-CSI-Feedback}. It is also necessary to evaluate the occupation of memory storage by NNs, which is majorly affected by the number of NN weights and the number of bits required to represent each NN weight. In practice, the UE-side AI/ML model needs to be designed as lightweight, given that UE has less computational power than gNB.

\emph{2) Model Generalization:} Whenever we train an AI model, we should pay attention to the model generalization \cite{Generalization}. Its enhancement is projected to be achieved through two key strategies: dataset mixing and online learning.

The model generalization error is usually caused by a mismatch in the distribution of training and test datasets. When an AI/ML model is trained using a mixed dataset compiled from multiple scenarios or configurations and subsequently subjected to inference or testing on a dataset from one of these specific scenarios/configurations, it results in a diminutive generalization error. This premise underpins the dataset mixing-based strategy. The process of selecting scenarios/configurations for validating generalization must take into account the inference nodes within the AI/ML model, such as gNB or UE, as well as the intended use cases, whether it be SBP or TBP. In practice, it is difficult for the mixed training dataset to cover all scenarios/configurations. Online learning is a good option to boost the generalization performance when a new scenario/configuration comes up. In fact, not only the air interface, but also the upper layer is involved in online learning, which requires a well-designed online learning protocol for the SBP and TBP. The protocol represents a significant challenge in the realm of AI-native air interfaces, necessitating clear guidelines for several critical aspects. These include determining the appropriate timing for conducting online learning, devising methods for the collection of BM samples, specifying which UEs should partake in the online learning process, and formulating strategies for the allocation of computing resources, such as power management. Meta and transfer learning \cite{Transfer-Learning} are key enablers for accelerating online learning and reducing training overhead.

Concurrently, the investigation and examination of generalization within AI-powered BM use cases must broaden its scope beyond the sole focus on a singular AI/ML model's adaptability to unfamiliar scenarios or configurations. It is imperative to entertain various other strategies such as developing a suite of specialized AI/ML models, each expertly attuned to distinct scenarios or configurations, with the flexibility to switch between these models depending on the specific deployment context. Moreover, for AI/ML models that operate on the UE side, there's a compelling need to delve deeper into how supplementary information influences their generalization capabilities. Signaling assistance information, for instance, could play a pivotal role in identifying scenarios and fostering model generalization, particularly by facilitating the strategic switching between models.

\begin{figure}[t]
	\centering
	\includegraphics[width=8.5cm]{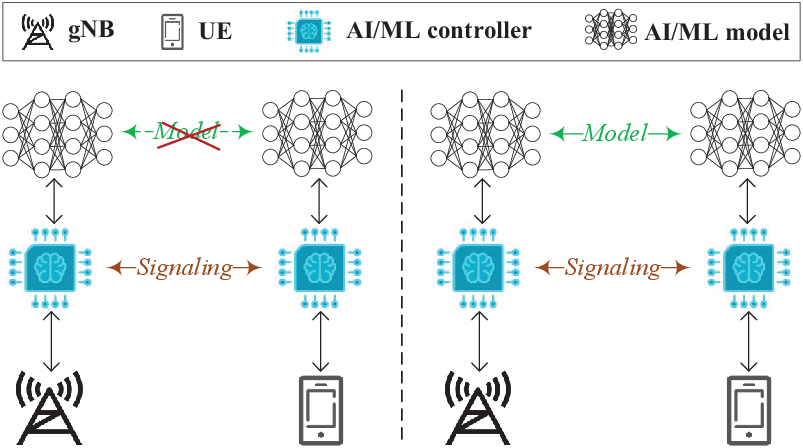}
	\caption{UE-gNB collaborations for AI-enabled BM.}
	\label{fig:collaboration}
\end{figure}

\emph{3) Collaboration between UE and gNB:} For the BM use cases leveraging AI/ML techniques, we identify and consider two types of UE-gNB collaboration shown in Fig.~\ref{fig:collaboration}.

$\bullet$ \emph{Type1: Collaboration without model transfer.} This type of collaboration mainly involves data collection and signaling-based AI/ML model management. UE may store several AI/ML models for different scenarios/configurations, and when the scenario/configuration changes, UE can select the proper one from these models. Accordingly, gNB needs to send some signals to the UE for model activation or switching. If the AI/ML models are not suitable for the current scenario/configuration, UE can collect training data on its own or through the network. Using the collected data, UE updates its AI/ML model to improve generalization performance. In addition, various kinds of AI/ML assistant information exchange may be considered, where signaling-based collaboration is also required.

$\bullet$ \emph{Type2: Collaboration with model transfer.} This collaboration type includes all collaboration operations in Type1. The difference lies in whether the AI/ML model can be transferred between gNB and UE. In this type, when the current model's performance declines, the model transfer is to enable the model switching (i.e., transfer the full model to replace the old one) or updating (i.e., reuse the model structure while updating only some model parameters). Taking into account the sensitivity of dataset privacy, the AI/ML model developed for BM may require training via approaches such as model splitting or federated learning \cite{Distributed-Learning} to preserve data confidentiality.

\emph{4) Model Life Cycle Management (LCM):} LCM is necessary for obtaining satisfactory BM performance by applying AI/ML algorithms.

$\bullet$ \emph{Data collection:} Given the purpose of data collection in LCM, reducing its overhead is critical while a large-scale and high-accuracy dataset is generally beneficial for improving model performance. On one hand, non-real-time and batch reporting may be needed to improve efficiency for purposes with less stringent latency requirements such as model training and updating. On the other hand, purposes with stricter latency requirements such as model inference and monitoring would require (near-)real-time measurement and reporting. In such a case, data compression may be necessary. Data quality could be ensured by selectively collecting/sending a subset of the overall collected data after data quality measurement processing, thereby reducing the overhead of data collection.

$\bullet$ \emph{Model training:} Based on the discussion on SBP and TBP, the AI/ML model training procedure can be completed by either UE or gNB, which should consider two aspects: data and computing power. The gNB has huge amounts of data in various conditions and owns strong computing power, which could help to develop an AI/ML model with good inference accuracy or good generalization capability. By contrast, UE's functional capabilities rely on its physical capabilities.

$\bullet$ \emph{Model inference:} The model training and inference location may be the same or different, which can impact the beam prediction solution. This discussion is related to the collaboration types presented above. In order to facilitate the inference, the study can start with enhanced or new beam measurement and/or beam reporting, enhanced or new signaling for measurement configuration/triggering, signaling of assistance information (if applicable), etc. For example, the gNB transmit beam pattern can be used to improve inference performance at UE. Also, the UE location and speed are useful for model inference in TBP.

$\bullet$ \emph{Model monitoring:} For BM, the beam prediction accuracy related KPIs (e.g., Top-$N/1$ beam prediction accuracy, L1-RSRP difference) and link quality related KPIs (e.g., throughput, L1-RSRP) are the preferred performance metrics of model monitoring. It is necessary to study the monitoring mechanisms used for UE/gNB-sided AI/ML models, whose design can focus on who monitors performance metric(s) and who makes decision(s) of model selection/activation/deactivation/switching/fallback operation. In addition, there are other key issues to be addressed, such as how often a model monitoring instance is needed? The option may be to execute periodically or after each prediction.

$\bullet$ \emph{Model transfer:} The two-side training and inference (e.g., model training at the gNB side and delivered to UE for inference or verse vice) would necessarily involve model transfer. It may face some restrictions in realistic networks. For example, gNB pre-trained model could be cell-specific and may face the generalization issue when serving different receive beam settings for different UEs. The UE pre-trained model could be UE-specific, which may not be applicable to another UE in the same cell. The capacity for model transfer should take into account the compatibility of supported structures, quantization, and processing capabilities between the UE and gNB.

\emph{5) Limitations of AI-enabled Framework:} While AI/ML-based BM is promising, it has limitations, including data dependency, necessitating large, high-quality datasets, which can be scarce. Non-stationary environments, where conditions change rapidly, can also impact AI/ML models' effectiveness, as they are typically trained on historical data. Addressing these challenges requires strategies such as data augmentation and transfer learning. Additionally, the use of AI/ML in BM may raise ethical and privacy concerns, especially when dealing with sensitive user data. To address this, techniques like differential privacy or federated learning can be employed. Ensuring the security of AI/ML models against adversarial attacks is also critical. Models must be robust to malicious inputs that could potentially lead to beam misalignment and performance degradation. Data encryption and access controls are key defenses.

\emph{6) Others:} During deployment, more issues and challenges must be overcome. For example, the wireless environment is dynamic, with changing user patterns, mobility, and physical obstructions. AI/ML models must continuously learn and adapt to these changes to maintain optimal BM performance. Achieving seamless integration of AI/ML-based systems with existing network management frameworks is essential. This requires careful design and validation to ensure network stability and reliability are maintained. Systems must be scalable to accommodate a large number of devices and the dynamic nature of wireless networks. This includes scalability in terms of algorithm complexity, processing power, and energy consumption. In real-time BM, AI/ML models must make predictions and decisions within strict latency requirements to avoid service degradation. This requires not only fast algorithms but also efficient hardware implementations.

\section{Conclusion}
This article presents two prominent AI-enabled BM frameworks: the SBP and TBP. It also explores a broader scope of AI/ML applications in BM and discusses open issues in the standardization aspects of AI/ML for BM, with a particular focus on critical elements such as complexity, generalization, collaboration, and LCM. This extensive exploration aims to inform and shape the future development of AI-enabled BM in 5G-Advanced and beyond.

\section*{Acknowledgment}
This work was supported in part by the National Natural Science Foundation of China under Grant U23A20279, Grant 62261160650, and Grant 62271094; in part by the Natural Science Foundation of Chongqing under Grant CSTB2024NSCQ-MSX0535; in part by the Science and Technology Research Program of Chongqing Municipal Education Commission under Grant KJQN202300646; in part by the Guangdong Basic and Applied Basic Research Foundation under Grant 2023A1515110732; in part by the Science and Technology Development Fund, Macau SAR under Grant 0087/2022/AFJ and Grant 001/2024/SKL; in part by the Research Committee of University of Macau under Grant MYRG-GRG2023-00116-FST-UMDF and Grant MYRG2020-00095-FST. Jiajia Guo and Shaodan Ma are co-corresponding authors of this article.

\ifCLASSOPTIONcaptionsoff
  \newpage
\fi

\bibliographystyle{IEEEtran}
\bibliography{reference}

\begin{IEEEbiography}[{\includegraphics[width=1in,height=1.25in,clip,keepaspectratio]{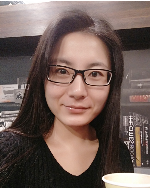}}]{Qing Xue}(xueq@cqupt.edu.cn) received the Ph.D. degree from Southwest Jiaotong University in 2018 and then joined the Chongqing University of Posts and Telecommunications. She was a post-doctoral fellow with the National Key Laboratory of Wireless Communications, University of Electronic Science and Technology of China. She was also a post-doctoral fellow with University of Macau under the Macao Young Scholars Program. Her research interests include millimeter-wave communications, intelligent wireless networking.
\end{IEEEbiography}
\begin{IEEEbiography}[{\includegraphics[width=1in,height=1.25in,clip,keepaspectratio]{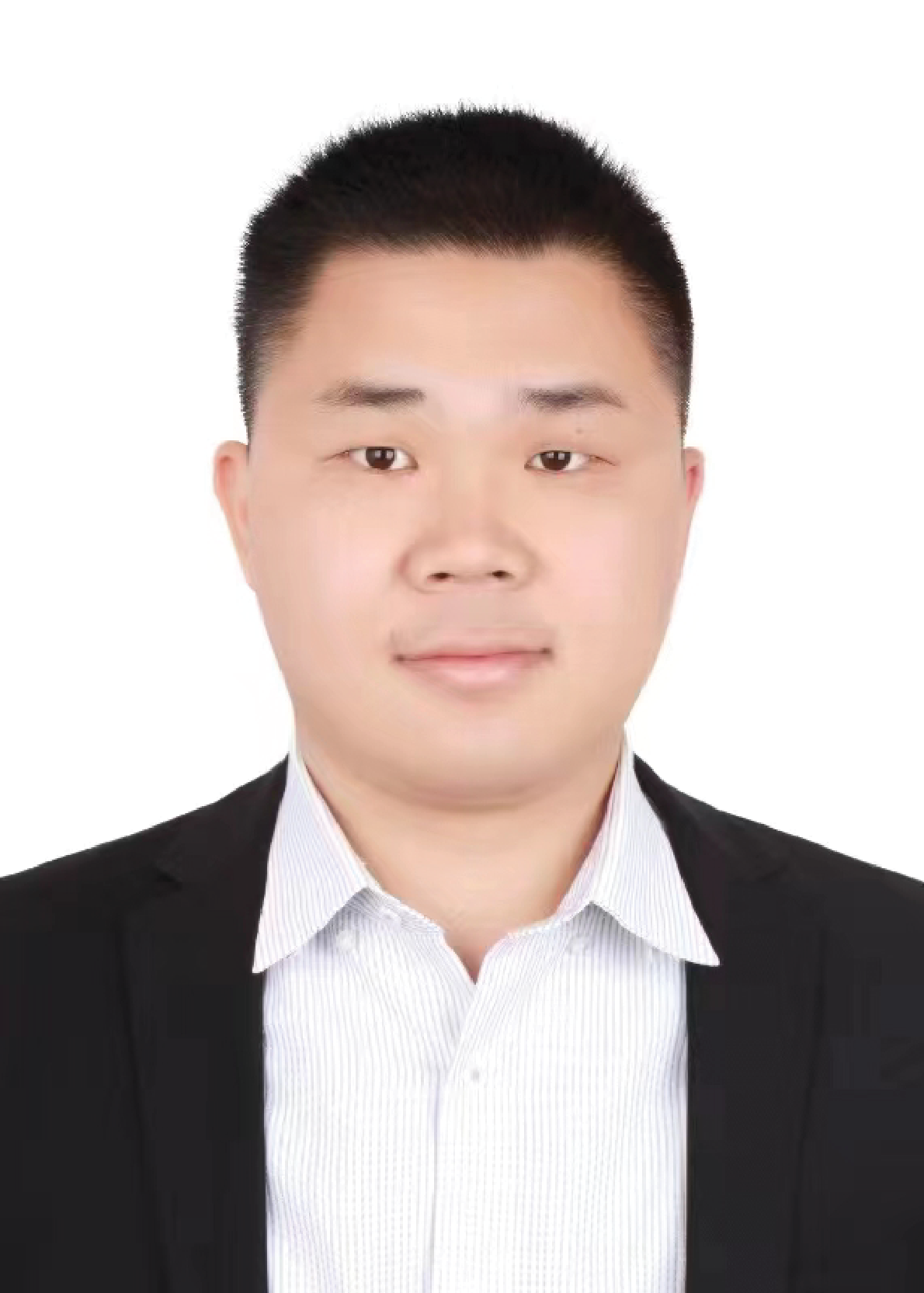}}]{Jiajia Guo}(jiajiaguo@seu.edu.cn) is with the State Key Laboratory of Internet of Things for Smart City, University of Macau, Taipa, Macau, P.R. China and also with the Zhuhai UM Science \& Technology Research Institute, Zhuhai, Guangdong, P.R. China. His current research interests include AI-native air interface, reconfigurable intelligent surfaces, and massive MIMO.
\end{IEEEbiography}
\begin{IEEEbiography}[{\includegraphics[width=1in,height=1.25in,clip,keepaspectratio]{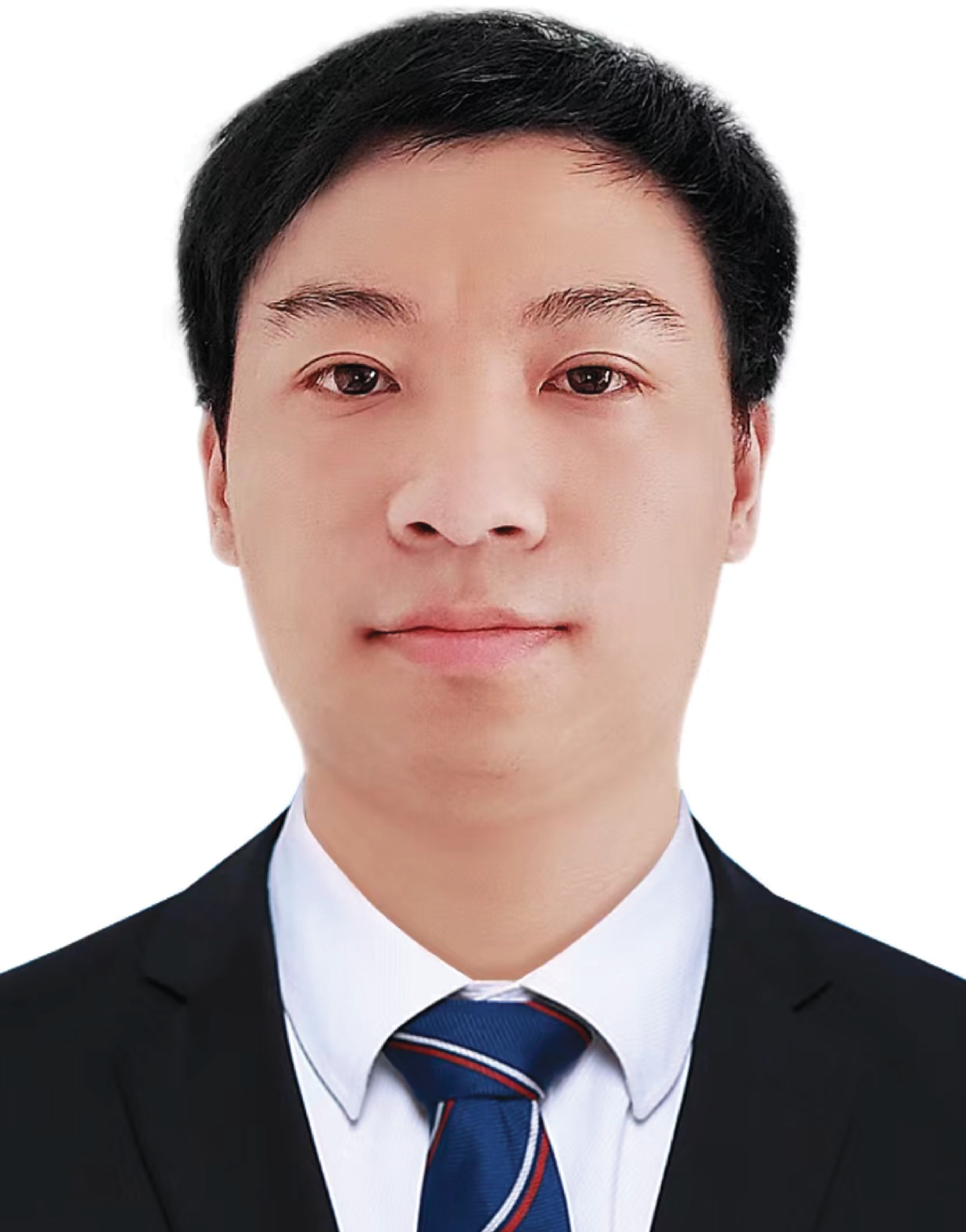}}]{Binggui Zhou} (binggui.zhou@connect.um.edu.mo) is currently working toward the Ph.D. degree in Electrical and Computer Engineering with the University of Macau, Macao, China. He also serves as a Research Assistant with the School of Intelligent Systems Science and Engineering, Jinan University, Zhuhai, China. His research interests include artificial intelligence (AI), AI empowered wireless communications, and data mining.
\end{IEEEbiography}
\begin{IEEEbiography}[{\includegraphics[width=1in,height=1.25in,clip,keepaspectratio]{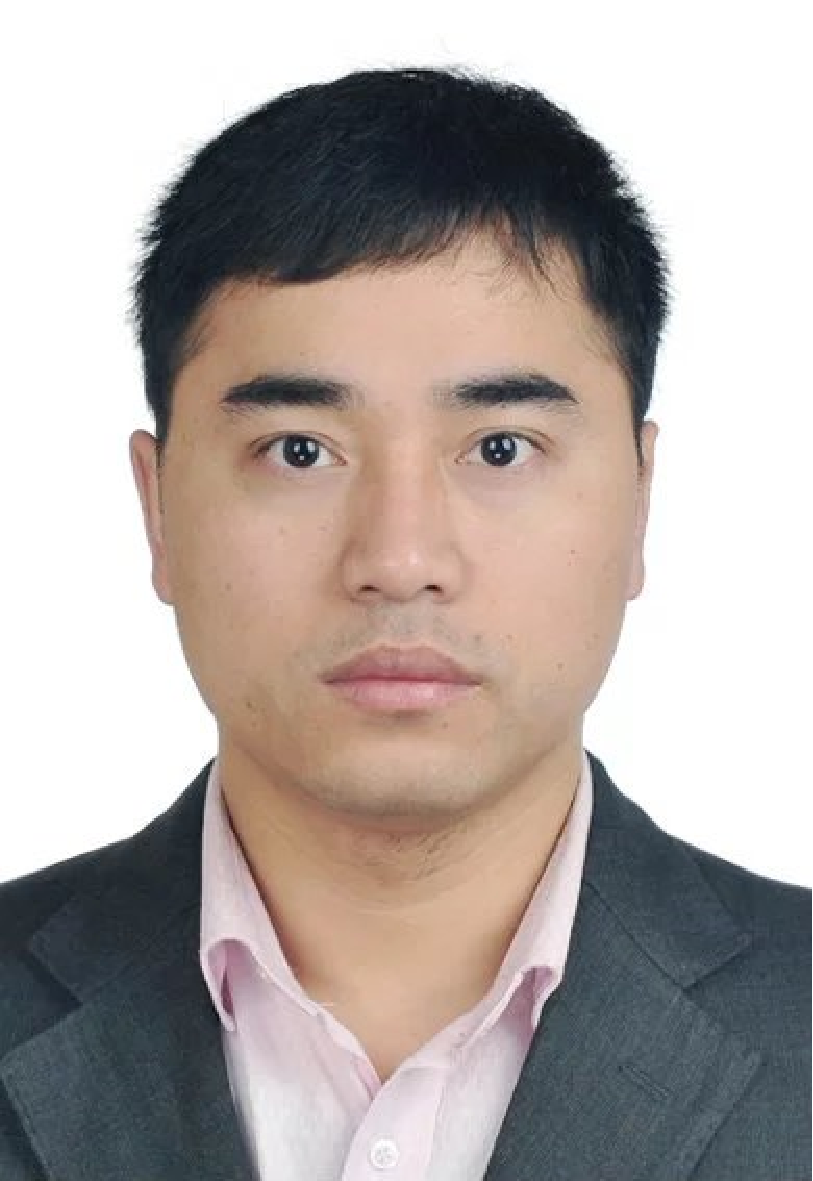}}]{Yongjun Xu}(xuyj@cqupt.edu.cn) is a Professor in CQUPT and was a Visiting Scholar in Utah State University. His recent interests include C-V2X, intelligent reflecting surface, etc. He has published more than 100 papers. He serves as an editor for several international journals. He serves on the Chair of ICCT 2023, ICCC 2020, WCSP 2021, and also a TPC member of many IEEE international conferences, such as Globecom, ICC, ICCT, WCNC, ICCC, and CITS.
\end{IEEEbiography}
\begin{IEEEbiography}[{\includegraphics[width=1in,height=1.25in,clip,keepaspectratio]{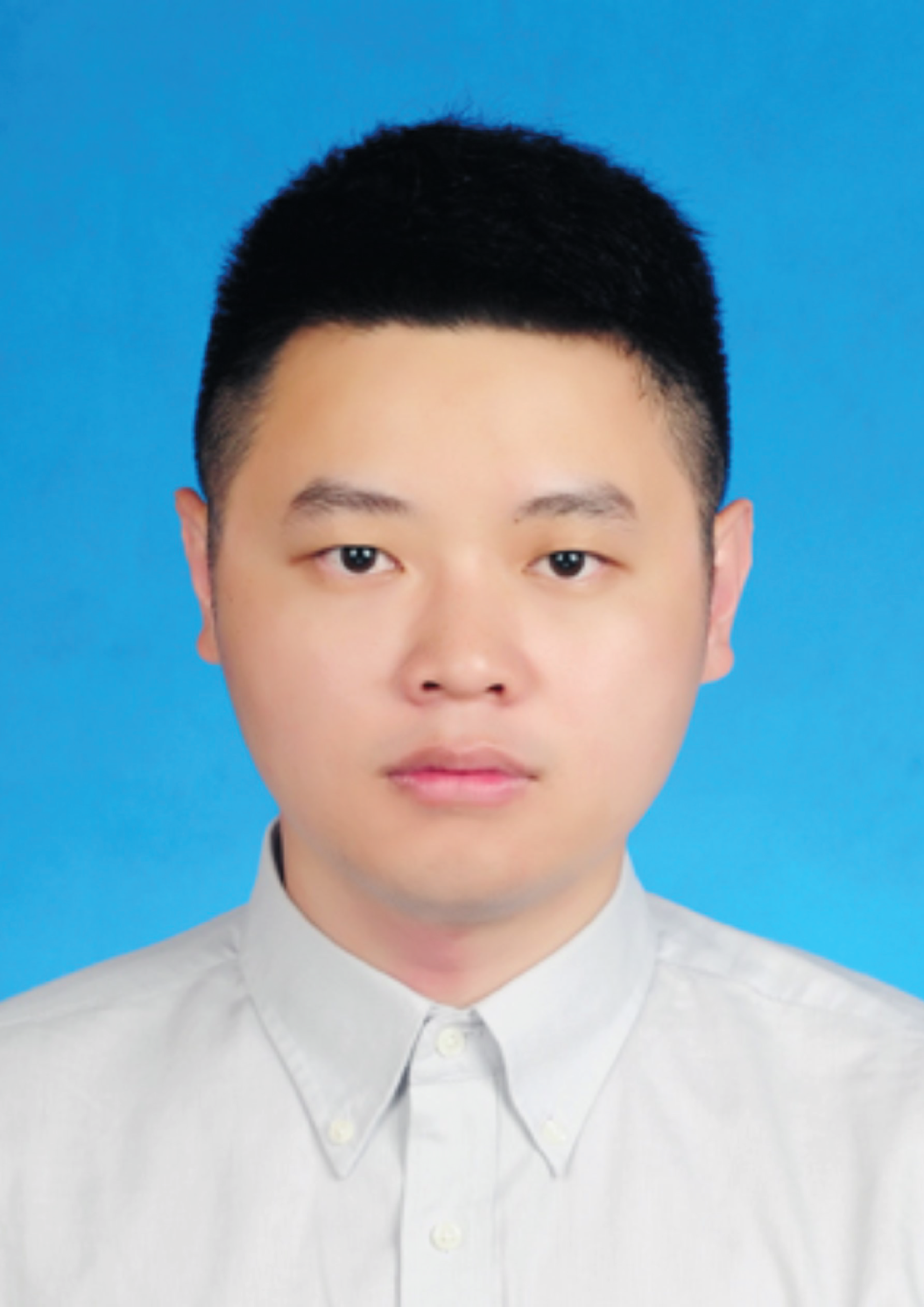}}]{Zhidu Li}(lizd@cqupt.edu.cn) received the PhD. degree in information and communications engineering from Beijing University of Posts and Telecommunications in 2018. He is currently a professor with Chongqing University of Posts and Telecommunications. His research interests include network calculus, wireless powered IoT, and edge intelligent.
\end{IEEEbiography}
\begin{IEEEbiography}[{\includegraphics[width=1in,height=1.25in,clip,keepaspectratio]{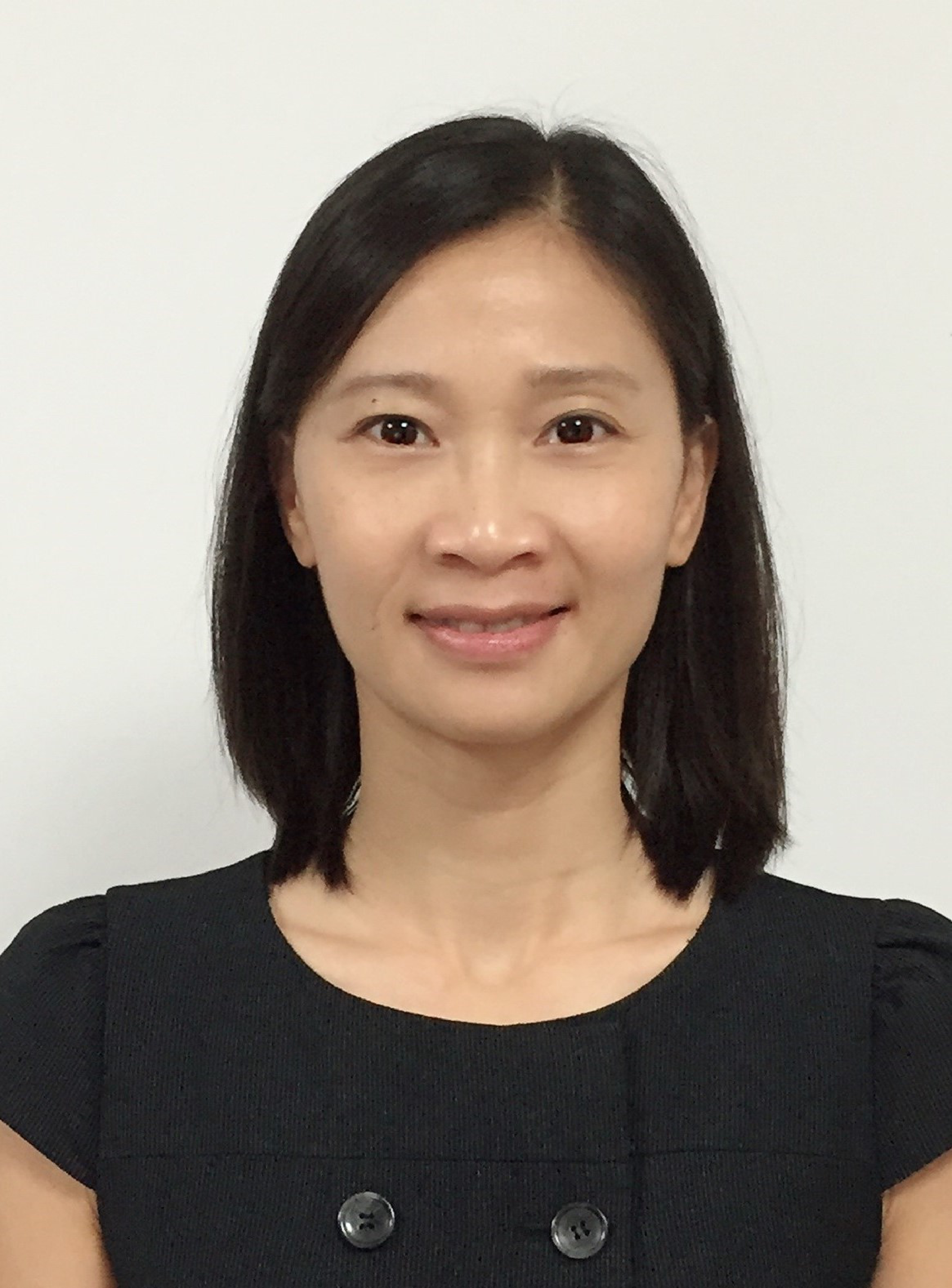}}]{Shaodan Ma}(shaodanma@um.edu.mo) is a Professor with the University of Macau. Her research interests include array signal processing, integrated sensing and communication, and massive MIMO. She was a symposium co-chair for various conferences including IEEE ICC 2021, 2019 \& 2016, IEEE/CIC ICCC 2019, IEEE GLOBECOM 2016, etc. She serves as an Editor for IEEE TWC (2018-2023), IEEE TCOM (2018-2023), IEEE WCL (2017-2022), IEEE CL (2023-present), etc.
\end{IEEEbiography}

\end{document}